\documentclass[aps, prl, reprint, showpacs, twocolumn, 10pt, groupedaddress]{revtex4-1}

\usepackage{graphicx, amsmath, dsfont, dcolumn, bm, times, color, braket, amssymb, multirow}

\begin{document}

\bibliographystyle{apsrev4-2}

\title{Effective increase of a superconducting critical temperature due to a configurational entropy of impurities}

\author{Viktoriia Kornich}
\affiliation{Institute for Theoretical Physics and Astrophysics, University of W\"urzburg, 97074 W\"urzburg, Germany }

\date{\today}

\begin{abstract}
We show theoretically that a superconducting critical temperature can be effectively increased in a fine-grained superconductor, where impurities migrate along the grain boundaries. The  distribution of impurities affects the spatial profile of the BCS condensate. Thus, BCS condensate acquires entropy equal to the configurational entropy of impurities. In order to employ this entropy into the superconducting phase dynamics, we suggest to use a quasiparticle trap that removes quasiparticle excitations from the superconductor. This makes the concentration of Cooper pairs a dynamic variable of the entropy, and thus influences the Ginzburg-Landau functional of the superconductor effectively reducing the first expansion coefficient or, in other words, increasing the critical temperature.\end{abstract}

\maketitle

\let\oldvec\vec
\renewcommand{\vec}[1]{\ensuremath{\boldsymbol{#1}}}
{\it Introduction.--} Superconductivity is a very diverse field including experimental and theoretical studies of bulk unconventional superconductors \cite{ran:science19, pustogow:nature19}, light-enhanced superconductivity \cite{mankowsky:prb15, michael:prb20}, acoustic-wave-induced superconductivity \cite{kornich:scipost22}, moir{\'e} materials \cite{kim:nature26}, van der Waals quasicrystals \cite{tokumoto:natcom24}, finite-energy superconductivity \cite{bahari:prr22}, finite-energy Cooper pairing \cite{tang:prl21, kornich:prb24}, and many others. The main criteria for superconductivity are negligibly low resistivity and a diamagnetic Meissner effect. Although, the last one is under scrutiny for the odd-frequency superconductors \cite{heid:zpb95, solenov:prb09}. The very low resistivity takes place in a wide variety of conventional and unconventional superconductors, however this usually does not happen at standard ambient conditions, e.g., the critical temperature of 250 K was reached in LaH$_{10}$ at 170 GPa \cite{drozdov:nature19}. There are various methods aimed at increasing superconducting critical temperature, including extraction of quasiparticles employing superconductor-insulator-superconductor junctions \cite{parmenter:prl61, chi:prb79}.

Entropy of electrons is usually low, and thus is rarely taken into account except, e.g., when the electron spectrum has high-entropy degeneracies \cite{aliani:prb26}. For a free non-interacting Fermi gas, entropy comes from the energy states with the occupation $<1$. Atomic configurational entropy is widely employed in the field of high-entropy alloys \cite{miracle:am17}. However, the role of configurational entropy for the superconducting critical temperature $T_c$ is unclear due to many material properties the entropy affects \cite{he:physrep25}.

\begin{figure}[b]
		\includegraphics[width=0.5\linewidth]{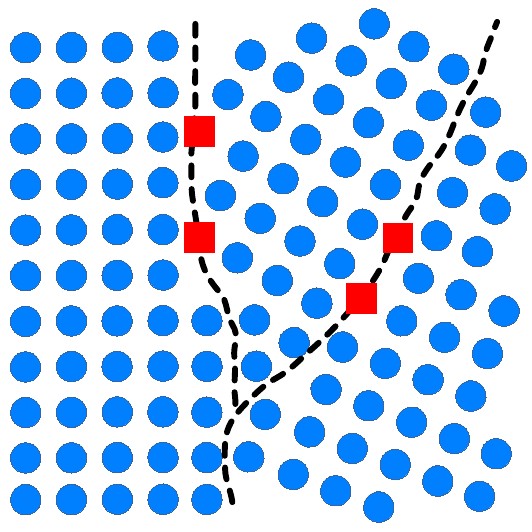}
		\caption{Schematic picture of a solid material (blue circles) consisting of grains with the grain boundaries shown by the black dashed lines. Impurities (red squares) can migrate along those boundaries due to non-zero temperature and quantum tunnelling effects. Their migration can be enhanced or suppressed by electric fields. }		\label{fig:setup}
\end{figure}

In this work, we suggest to consider a crystal with a fine-grained structure \cite{kiana:prm24}. Here, we will study the case, when their presence does not destroy the superconductivity of the sample, but consider the grain boundaries as channels for impurities to move around the crystal \cite{kim:jap13}, see Fig.~\ref{fig:setup}. While weak disorder with charge impurities should not destroy the superconductivity, charge impurities affect the spatial distribution of the wavefunction of the BCS condensate \cite{ghosal:prl98}. Thus, each configuration of disorder gives a different spatial profile of a BCS condensate wavefunction. Therefore, even though the BCS condensate is described via a single wavefunction, it has entropy that is equal to configurational entropy of impurities. This entropy depends on the structural properties of the crystal, therefore it gives a mere shift of the free energy of the superconductor, but does not affect the superconducting transition. 

In order to involve the entropy into the superconducting phase space definition, we suggest to use a quasiparticle trap, that absorbs quasiparticle excitations from the sample \cite{parmenter:prl61, chi:prb79, goldie:prl90, ullom:prb00, rajauria:prb12, knowles:apl12, wang:natcom14, riwar:prb19, marin:nanolett20} formed by $N_q$ electrons. In such a way, we make a crystal positively charged. The effect of charging of a sample can vary from negligible to drastic depending on the material type, method of charging, and the dimensionality of the sample. In this work, we do not focus on a certain material and a sample form. We suggest that this material has grain structure and the induced positive charge leads to the new charge distribution around the boundaries of the grains, which can lead to enhanced or suppressed impurity migration. If impurities are positively charged, they will have less sites to populate, which we account for as $\eta N_q$, where $-1<\eta<0$.


In such a way, the number of Cooper pairs becomes a variable of the entropy, because $N_q=N_c-N_s-N_n$, where $N_c$, $N_s$, and $N_n$ are the numbers of all conducting electrons, superconducting electrons, and low-energy conducting normal electrons, respectively. Now, the entropy contributes to the Ginzburg-Landau functional not as a mere shift in energy, but changing its shape and its extrema. Namely, the contribution of the entropy enters the first expansion coefficient in $|\Delta|^2$, where $\Delta$ is the superconducting order parameter, effectively lowering the actual temperature of the sample, $T$, and thus effectively increasing the critical temperature. We discuss the parameters of the sample, that can make this correction crucial. We then study the process of equilibration of superconductivity, as the quasiparticle trap works on the finite time scales, and during this time quasiparticles can recombine into Cooper pairs. Then we show that the spectral gap of the Higgs modes can serve as a direct probe of the discussed effect. We finish by mentioning other physical quantities, such as London penetration depth, that can be helpful in detection of this effect. Importantly, our consideration is very general: Ginzburg-Landau theory and phenomenology. Therefore, this effect can potentially take place in many materials.

 {\it The origin of entropy.--} We consider a conducting material consisting of grains that turns into a superconductor at sufficiently low temperatures. Its grain structure does not destroy superconductivity, but provides channels for impurities to migrate \cite{kim:jap13}. Impurities can move between some $N$ low-energy sites (e.g., in the atomic vacancies at the grain boundaries) in these channels due to finite temperature and quantum tunnelling. If the amount of impurities is sufficiently low compared to $N$, we can neglect impurity-impurity interactions.  Thus, after enough long time it does not matter, what was the initial distribution of impurities between those low-energy sites, because on a macroscopic scale, it will eventually become random and will change randomly.
 
 If we have $N_i$ amount of impurities, and they are randomly distributed among $N$ low-energy sites, we obtain an entropy of impurities,
 \begin{eqnarray}
 \label{eq:Stilde}
 \tilde{S}_i=k_B \ln{\left[\frac{N!}{N_i! (N-N_i)!}\right]},
 \end{eqnarray}
 analogously to the entropy of holes migrating between possible sites \cite{schleger:prb94, tetot:prb99, zhou:prl06}.
 
Now, let's assume, we removed $N_q$ electrons from $N_c$ conducting electrons of the crystal. This means we eliminated its electroneutrality: it is charged positively now and the potential landscape where impurities are moving has been changed, e.g., the number of sites, where they can reside changed by $\eta N_q$, where $|\eta|<1$ is a number. As $N_q\ll N_c$, we take only the linear part of this effect, $\propto \eta$.

Then the entropy of impurities is, 
\begin{eqnarray}
S_i=k_B \ln{\left[\frac{(N+\eta N_q)!}{N_i! (N+\eta N_q-N_i)!}\right]}.
\end{eqnarray}
If we use Stirling's formula we obtain,
\begin{eqnarray}
&&S_i\simeq-k_B\left(N_i\ln{\frac{N_i}{N+\eta N_q}}+\right.\\ &&\left.+(N+\eta N_q-N_i)\ln{\frac{N+\eta N_q-N_i}{N+\eta N_q}}\right).
\end{eqnarray} 
If we express $N_q$ electrons as $N_q=N_c-N_s-N_n$, meaning that $N_q$ is the electron number that contribute to Bogoliubov quasiparticles, $N_s$ denote superconducting electrons constituting superconducting condensate, and $N_n$ electrons are the rest of normal electrons up to the lower bound of the conduction band. Then, assuming $|\eta| N_s\ll N+\eta(N_c-N_n)-N_i$ the part of the entropy with $N_s$ is
\begin{eqnarray}
\label{eq:Ss}
\nonumber S_s=k_B N_s\eta\ln{\frac{N-N_i+\eta(N_c-N_n)}{N+\eta(N_c-N_n)}}-\\ -\frac{k_B}{2}N_s^2\frac{\eta^2N_i}{(N+\eta(N_c-N_n))(N-N_i+\eta(N_c-N_n))}=\nonumber \\ =k_B \Sigma_1N_s-\frac{k_B}{2}\Sigma_2N_s^2.\ \ \ \ 
\end{eqnarray} 
The prefactor before $N_s$ is still positive, when
\begin{eqnarray}
-\frac{N-N_i}{N_c-N_n}<\eta<0,
\end{eqnarray} 
meaning that the amount of possible sites must be reduced, but stay within the certain range.

{\it The main statement.--} If the resulting material is in a superconducting state, the superconducting electrons $N_s=N_c-N_n-N_q$ form $N_s/2$ Cooper pairs. The Ginzburg-Landau functional, $\mathcal{F}=\int Fd{\bm r}$, has the following density,
\begin{eqnarray}
\label{eq:Fin}
F=F_0+2\alpha |\Delta|^2+2\beta|\Delta|^4-T\frac{S_i}{V},
\end{eqnarray} 
where $F_0$ is the non-superconducting part of the functional for the conducting electrons, $V$ is the volume of the whole sample, $\Delta$ is the order parameter, $\alpha$ and $\beta$ are Ginzburg-Landau coefficients. Importantly, the concentration of Cooper pairs, i.e., pairs of superconducting electrons is $n_s/2=|\Delta|^2$ \cite{abrikosov:book17}, where $n_s=N_s/V$. Analogously, $n_n=N_n/V$, $n_q=N_q/V$, $n_c=N_c/V$ are the concentrations of different normal and conducting electrons, respectively.

Here, we use $S_i$ as an entropy of a superconducting condensate, because different distribution of impurities in a crystal means different spatial distribution of electrons that constitute the BCS condensate \cite{ghosal:prl98}. While the difference is likely to be not noticeable on the scale of the superconducting gap size, the BCS state will have different spatial profiles, meaning that there is an entropy due to all these possible states. 

We remove $N_q$ electrons via trapping of excited quasiparticles. The discussed quasiparticles are excitations of a superconductor, that is electron-hole Bogoliubov quasiparticles that diagonalize the BCS Hamiltonian \cite{abrikosov:book17}. The normal electrons, $N_q$, constitute quasiparticles. There are various methods of removing quasiparticles from a superconductor or its certain region, e.g., via a metallic trap, modulation of the superconducting gap, trapping into vortices \cite{goldie:prl90, ullom:prb00, rajauria:prb12, knowles:apl12, riwar:prb19, wang:natcom14} or within the superconducting junctions \cite{parmenter:prl61, chi:prb79, marin:nanolett20}. In order to have a stable state in the system consisting of the crystal and the trap, extracted electrons should have sufficiently low energy of the resulting state in the trap or the outer circuit attached to it. 

The Ginzburg-Landau coefficients $\alpha$ and $\beta$ depend on $T$ in general. We assume that $\alpha$ is a monotonic growing function of $T$ and must change its sign after the superconducting transition. Coefficient $\beta$ is usually taken to be positive, $\beta>0$. For conventional superconductors in the limit $T_c-T\ll T_c$, $\alpha=a(T-T_c)$, where $T_c$ is the critical superconducting temperature (in our case, the critical temperature of the full crystal consisting of grains and including impurities, but with the quasiparticle trap turned off), $a(T)=const>0$ \cite{abrikosov:book17}. The terms of $F$, containing $n_s$ are,
\begin{eqnarray}
\label{eq:Fs}
F_s=\left[\alpha-k_BT\Sigma_1\right]n_s+\frac{1}{2}\left[\beta+k_BTV\Sigma_2\right]n_s^2.\ \
\end{eqnarray}
We denote the term in square brackets with $\alpha$ as~$\tilde{\alpha}$ and the term in square brackets with $\beta$ as $\tilde{\beta}$. We can see that the term $\Sigma_1>0$ from the entropy keeps $\tilde{\alpha}<\alpha$, therefore the expected flip of sign of $\tilde{\alpha}$ will happen later than for $\alpha$. As a consequence, the equilibrium value $n_s=-\tilde{\alpha}/\tilde{\beta}\neq 0$ will hold for higher temperatures than $-\alpha/\beta$, thus effectively increasing $T_c$, even though $\tilde{\beta}>\beta$. 

In Eq.~(\ref{eq:Fs}), there is no an obvious limit, that would ensure the change of sign of $\tilde{\alpha}$. If we consider the linear expansion of $\alpha$ in the limit $T_c-T\ll T_c$, we will see that 
\begin{eqnarray}
\tilde{\alpha}=\left(a-k_B\Sigma_1\right)T-aT_c,
\end{eqnarray}
and there is no mechanism to guarantee that $a$ is always larger than the correction $k_B\Sigma_1$. This can potentially lead to an always present minimum $n_s=-\tilde{\alpha}/\tilde{\beta}$. However, as we will discuss later, quasiparticle traps also have their limitations, therefore at an enough high $T$, Eq.~(\ref{eq:Fs}) stops working. However, we think that in the majority of cases, the coefficient $\alpha$ should overpower the correction at some point, and this should come naturally from the values of parameters of the system. 

Important question is whether $\alpha$ is by order of magnitude comparable to the correction induced by the entropy. Let's evaluate it. Density of a free energy of a superconductor of type I equals to the density of energy of an electromagnetic field that takes the volume of the superconductor after the transition, i.e., $F_H=\mu_0 H_c^2/2$, where $\mu_0$ is a vacuum permeability and $H_c$ is a critical magnetic field. Let's take $\mu_0 H_c=0.01$ T, then $F_H\approx40$ J/m$^3$. 

Let's evaluate the order of the first term of Ginzburg-Landau expansion as $F_H\sim a(T_c-T) n_s$. If we take $T_c-T=0.1$ K and $n_s\simeq 1.5\times 10^{26}$ m$^{-3}$, we obtain $a\approx 0.19 k_B$. 

Now, let's reason the value of $n_s$. Let's take the concentration of the conducting electrons as $n_c=10^{29}$ m$^{-3}$. We evaluate the concentration of electrons that are affected by superconductivity by $n_s+n_q=4\pi p_F^2\Delta p/(2\pi \hbar)^3$, where $p_F=\hbar(3\pi^2n_c)^{1/3}$, $\Delta p\simeq 0.01 p_F$ and obtain $n_s+n_q=1.5\times 10^{27}$ m$^{-3}$. Taking into account that we consider $T-T_c\ll T_c$, we take $n_s=0.1(n_s+n_q)=1.5\times 10^{26}$ m$^{-3}$.

Now, if we take the concentration of sites available for impurities $N/V=10^{26}$ m$^{-3}$ and the amount of impurities $N_i=0.1 N$, and $\eta=-0.059$ (the lower limit for $\eta$ is $-(N-N_i)/(N_c-N_n)=-0.06$) the correction is $\Sigma_1=0.12$. Thus, $a/(k_B\Sigma_1)\approx 1.6$, so the effect of the correction is crucial. However, if we take $\eta=-0.04$, then $\Sigma_1\approx 0.012$, so in this case, the correction can be noticeable, but not crucial. Therefore, it requires certain fine tuning of parameters in order to make the obtained correction $k_B\Sigma_1$ significant. 

The correction term to $\beta$, i.e., $k_BTV\Sigma_2$, has the order of $\sim k_BT\eta^2/n_e$, where $n_e$ is the quantity with the units of concentration of electrons and is of the order of $10^{24}$ m$^{-3}$ for the parameters shown above. Therefore, we will omit $k_BTV\Sigma_2$ from $\tilde{\beta}$ and use just $\beta$ in the following.

{\it Quasiparticle trap.--} While the current work does not focus on the functioning details of quasiparticle traps, we will shortly mention different options. There are various types of quasiparticle traps. If the studied material is a superconductor of type II, trapping in vortices can be useful \cite{wang:natcom14}. Traps based on the N-I-S contacts \cite{artemenko:ssc79, blonder:prb82} and gap modulation \cite{aumentado:prl04, yamamoto:apl06} are rather popular in the field of superconducting qubits \cite{riwar:prb16, hosseinkhani:prb18, riwar:prb19, schmit:arxiv20}, however, they are usually engineered to work at very low temperatures, $T\ll T_c$. While Ginzburg-Landau theory is used also far from the superconducting transition nowadays, this is not a strictly-reasoned approach yet. There is also a method of quasiparticle trapping via superconducting junctions \cite{parmenter:prl61, chi:prb79, marin:nanolett20}, which initially was proposed to increase a superconducting critical temperature, and the experimental study reached temperatures close to $T_c$ \cite{chi:prb79}. Although, the functioning of this trapping is limited by at least the critical temperatures of the constituent parts of the junction.

To proceed with our study, we assume that the density of quasiparticles is decaying proportional to $e^{-\Gamma t}$, where $\Gamma$ contains different physical processes and rates present in the system, i.e., tunnelling into the neighbouring superconductor or metal, escape back into the sample under study, relaxation to the lower energy state, etc. We will use this exponential decay in our investigation of equilibration of superconductivity.

\begin{figure}[b]
		\includegraphics[width=\linewidth]{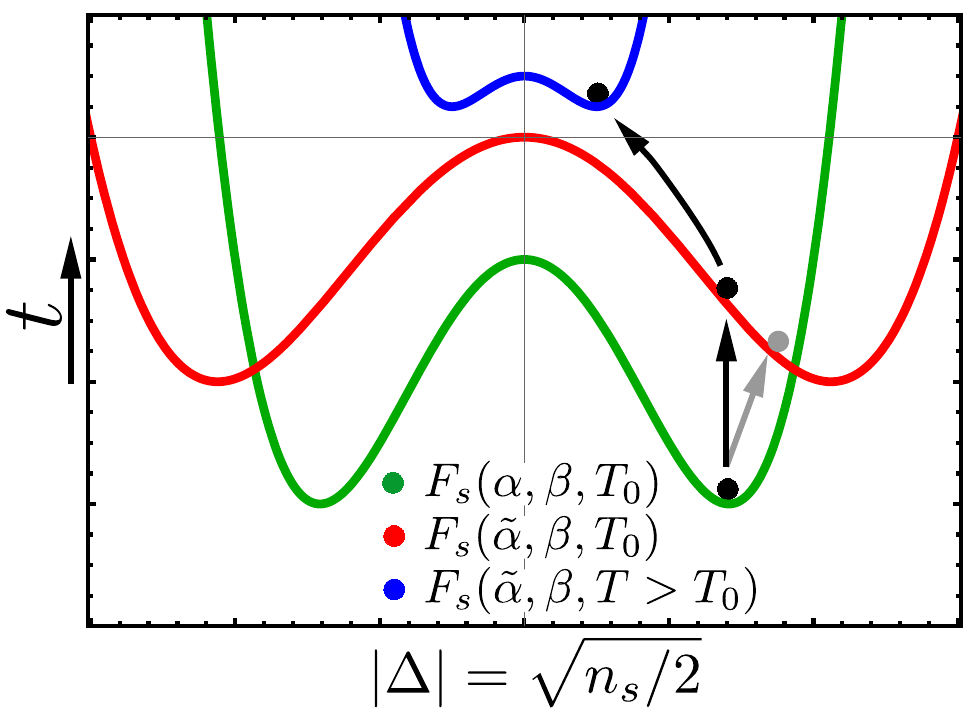}
		\caption{Evolution of the Ginzburg-Landau functional $F_s$ and the free energy of the superconducting sample in time $t$. The evolution in time of the free energy (black circles) is noted via black arrows. The green line shows the initial Ginzburg-Landau functional of the sample at a certain temperature $T_0$, described by $\alpha$ and $\beta$; so far $S_i(\eta=0)=const$. Then we turn on a quasiparticle trap, that absorbs quasiparticles and in such a way normal electrons $N_q$. The decrease of energy of electrons in a quasiparticle trap or in the outer circuit attached to it must be sufficient for this process to be effective. The functional $F_s$ is now described by $|\tilde{\alpha}|>|\alpha|$ (red line) because the entropy also contributes to it, however, as Cooper pairs are kept in the sample, $n_s$ value stays approximately constant (see the position of the black circle). The correction due to a finite function time of the trap can slightly shift $n_s$ to a larger value (grey circle, grey arrow). The temperature increases further, while the trap is kept connected. At some point, $n_s$ becomes the minimum of $F_s$ and since then stays in the minimum, while temperature grows and $n_s$ decreases (blue). In the first regime (green), the Higgs modes have the gap $-4\alpha$, while in the last one (blue), $-4\tilde{\alpha}$, see Eq.~(\ref{eq:FHiggs}).}		
		\label{fig:FreeEnergy}
\end{figure} 

{\it Equilibration of superconductivity.--} Here, we will study, how the concentration of Cooper pairs, $n_s$, equilibrate after the trap starts working, e.g., once the barrier between the sample and the trap is lowered or thinned or the bias is turned on. As $n_q$ decays, the influence of the correction from the entropy on the coefficient $\alpha$ grows. Therefore, some of the normal electrons $\delta n_q$ (while there is still finite $n_q\neq 0$ in the crystal) can recombine into Cooper pairs, thus decreasing the energy of the sample. If this happens, the final value of $n_s$ will not be precisely $|\alpha|/\beta$, but somewhat larger, see the grey circle and the grey arrow in Fig.~\ref{fig:FreeEnergy}. We denote the recombination rate of normal electrons into Cooper pairs due to this process with $\gamma$. 
In this section, we neglect the possibility for low-lying $N_n$ electrons to be excited onto the empty electron states within the energy range involved in superconductivity, $E-E_F\in [-k_B\Theta, k_B\Theta]$ with $E_F$ being Fermi energy (in case of a conventional superconductor, $\Theta$ is a Debye temperature). In principle, there can be thermal excitation of low-lying electrons to higher energy levels via phonons, but it will not be noticeable, if $T_c\ll \Theta$. For Al, for example, $\Theta\approx 428$ K \cite{phillips:pr59} and $T_c\approx 1.2$ K \cite{caplan:pr65}. 

In order to evaluate the effect of equilibration described above, we cannot use the time-dependent Ginzburg-Landau theory due to a rather long $1/\Gamma$ \cite{kopnin:book09} that can in principle be $\sim 10^{-3}$ s \cite{riwar:prb16}. Therefore, we will use a phenomenological description,
\begin{eqnarray}
\label{eq:nndiff}
\frac{dn_q}{dt}&=&-\Gamma n_q-\gamma n_q^2,\\
\label{eq:nsdiff}
\frac{dn_s}{dt}&=&\gamma n_q^2,
\end{eqnarray} 
where $t$ is time; $t=0$ denotes the start, when the trap is turned on, consequently $\gamma(t=0)=0$. The initial values are $n_s^0=|\alpha|/\beta$ and $n_q^0=n_c-n_s^0-n_n$. Eq.~(\ref{eq:nndiff}) is a Bernoulli equation that can be simplified with a substitution $n_q=1/y$. If we assume that the time dependence of $\gamma$ is enough slow, so that $d\gamma/dt \ll \Gamma^2 /n_q^0$,
\begin{eqnarray}
n_q=\frac{n_q^0 \Gamma}{\Gamma e^{\Gamma t}+\gamma n_q^0(e^{\Gamma t}-1)}.
\end{eqnarray}
Then $n_s$ can be obtained from Eq.~(\ref{eq:nsdiff}) by integration over time. If we neglect the time dependence of $\gamma$ and assume $\gamma\neq 0$, which is a valid approximation except for $t\approx 0$, we obtain 
\begin{eqnarray}
\nonumber n_s=n_s^0+n_q^0\frac{(\gamma n_q^0+\Gamma)(e^{\Gamma t}-1)}{\Gamma e^{\Gamma t}+\gamma n_q^0(e^{\Gamma t}-1)}+\nonumber\\ +\frac{\Gamma}{\gamma}\ln{\left(\frac{\Gamma e^{\Gamma t}}{\Gamma e^{\Gamma t}+\gamma n_q^0(e^{\Gamma t}-1)}\right)}.
\end{eqnarray}
In the end, the equilibrium value, i.e., at $t\rightarrow \infty$, is
\begin{eqnarray}
n_s=n_s^0+n_q^0+\frac{\Gamma}{\gamma}\ln{\left(\frac{\Gamma}{\Gamma+\gamma n_q^0}\right)}.
\end{eqnarray}
In the limit, $\gamma n_q^0\ll \Gamma$, the correction becomes just $n_s-n_s^0=\gamma (n_q^0)^2/(2\Gamma)$.

{\it Detection of the correction to $\alpha$.--} Let us suggest the scheme for detection and measurement of the correction to the coefficient $\alpha$. The gap in the spectrum of the Higgs modes should be proportional to $\tilde{\alpha}$, when we are in the regime of Eq.~(\ref{eq:Fs}) \cite{shimano:arcmp20}. Its value can be compared to the initial gap proportional to $\alpha$, when the quasiparticle trap was disconnected from the sample. 

We will follow the standard procedure for the derivation of the spectrum of the Higgs modes \cite{shimano:arcmp20}: We will take $F_s$ from Eq.~(\ref{eq:Fs}) (with $\tilde{\beta}\approx\beta$), add a kinetic term and the energy of the fluctuations of an electromagnetic field present in our sample, obtaining,
\begin{eqnarray}
\nonumber \mathcal{F}_s&=&\int d{\bf r}\left[2\tilde{\alpha} |\Delta|^2+2\beta |\Delta|^4+\right.\\ &&\left.+\frac{1}{2m}\Big|\left(-i\hbar\boldsymbol{\nabla}-\frac{q}{c}{\bf A}\right)\Delta\Big|^2+\frac{{\bf B}^2}{8\pi}\right],\ \ \ \ \ \
\end{eqnarray}
where $m$ is the mass of a Cooper pair, $q$ is the charge of a Cooper pair, $c$ is the speed of light, ${\bf A}$ is a vector potential, and ${\bf B}$ is the magnetic field, ${\bf B}=\boldsymbol{\nabla}\times {\bf A}$. We then expand $\mathcal{F}_s$ in terms of fluctuations up to the second order. The Higgs modes are the fluctuations of the amplitude of the order parameter, $\Delta=(\Delta_0+\delta\Delta)e^{i\phi}$, i.e., $\delta\Delta$. We obtain,
\begin{eqnarray}
\label{eq:tildeFfull}
\nonumber \mathcal{F}_s=\int d{\bf r}\left[2\Delta_0^2(\tilde{\alpha}+\beta\Delta_0^2)+2(\tilde{\alpha}+6\beta\Delta_0^2)\delta\Delta^2+\right.\\ \left.+\frac{\hbar^2}{2m}(\boldsymbol{\nabla}\delta\Delta)^2+\frac{\Delta_0^2}{2m}\Big|\hbar\boldsymbol{\nabla} \phi-\frac{q}{c}{\bf A}\Big|^2+\frac{{\bf B}^2}{8\pi}\right].\ \ \ \ \ \
\end{eqnarray}
The last two terms are usually considered in order to derive Anderson-Higgs mechanism. If $\Delta_0=\sqrt{-\tilde{\alpha}/(2\beta)}$, we obtain the spectrum of the Higgs modes via a Fourier transformation of the second and the third terms from Eq.~(\ref{eq:tildeFfull}), 
\begin{eqnarray}
\label{eq:FHiggs}
\mathcal{F}_{\rm Higgs}=V\int \left[-4\tilde{\alpha}+\frac{\hbar^2 {\bf k}^2}{2m}\right]\delta\Delta_{\bf k}\delta\Delta_{-{\bf k}}\frac{d{\bf k}}{(2\pi)^3}.
\end{eqnarray}
We can see that there is a gap, $-4\tilde{\alpha}$. Thus, this is the energy required to induce Higgs modes around $\Delta_0$. 

However, if we are in the regime shown by the red line and the black circle on it in Fig.~\ref{fig:FreeEnergy}, where the system is kept at $\Delta_0=\sqrt{-\alpha/(2\beta)}$, but the first Ginzburg-Landau coefficient is already $\tilde{\alpha}$, the calculation should include the reason, i.e., the forces, that keep the number of Cooper pairs constant. These forces are the ones that keep electrons bound to the crystal and repel others from the outside of it, so they are not included into Eq.~(\ref{eq:tildeFfull}), and the result of the measurement of these fluctuations $\delta\Delta$ will not directly inform us about the superconducting properties of the sample.

{\it Other types of measurement of $\tilde{\alpha}$.--} In the regime, shown by the blue line in Fig.~\ref{fig:FreeEnergy}, there are other quantities that include parameter $\tilde{\alpha}$, e.g., a London penetration depth $\lambda_L$ or a critical magnetic field $H_c$ \cite{abrikosov:book17}:
\begin{eqnarray}
\label{eq:lambda}
\lambda_L&=&\sqrt{\frac{mc^2}{2\pi q^2 n_s}}=\sqrt{\frac{mc^2\beta}{2\pi q^2 |\tilde{\alpha}|}},\\
H_c&=&2\sqrt{\frac{\pi \tilde{\alpha}^2}{\beta}},
\end{eqnarray}
respectively. However, as it can be problematic to approach the full destruction of superconductivity while keeping Eq.~(\ref{eq:Fs}) valid, it is better to use a superconductor of type II with $H_{c1}$ instead of a superconductor of type I with $H_c$. The critical magnetic field $H_{c1}$ is related to $H_c$ as \cite{abrikosov:book17, coleman:book15},
\begin{eqnarray}
H_{c1}\sim\frac{H_c}{\sqrt{2}}\frac{\ln{\kappa}}{\kappa}, \ \ \kappa\gg 1,
\end{eqnarray}  
where $\kappa=\lambda_L/\xi$ with $\xi=\xi_0\left(1-\frac{T}{T_c}\right)^{-1/2}$ being the Ginzburg-Landau correlation length.

{\it Conclusions.--} In this work, we consider a fine-grained superconductor, where the grain boundaries serve as channels for migration of charged impurities. The presence of impurities along the boundaries affects the shape of the wavefunction of a BCS condensate. Thus, every distribution of impurities has the corresponding BCS condensate wavefunction. Therefore, BCS condensate acquires entropy that corresponds to the configurational entropy of migrating impurities. However, this entropy, as it is, has no influence on a superconducting transition, because it is defined by the fabrication of the sample. In order to employ it, we suggest to use a quasiparticle trap that absorbs quasiparticles from this superconductor, thus making the concentration of Cooper pairs a dynamic variable of the entropy. The obtained contribution from the entropy decreases the first Ginzburg-Landau coefficient effectively increasing the critical temperature. 

 \begin{acknowledgments}
	I acknowledge useful discussions with Sergey N. Artemenko, Konstantin Nesterov, Atanu Patra, and Bj\"orn Trauzettel. This work was supported by the Würzburg-Dresden Cluster of Excellence ctd.qmat, EXC2147, project-id 390858490, and the DFG (SFB 1170).  \end{acknowledgments}

\end{document}